\documentclass[twocolumn]{aastex702}
\usepackage{amsmath}

\begin{document}

\title{ Diocotron Modes in Pulsar Magnetospheres: Charge Diffusion and Implications for Radio Emission Variability }

\author[0000-0003-1199-5552]{Matthew Goodbred}
\affiliation{Department of Astrophysical Sciences, Princeton University, 4 Ivy Lane, Princeton, NJ 08540, USA}
\email{mgoodbred@princeton.edu}
\author[0000-0001-9179-9054]{Anatoly Spitkovsky}
\affiliation{Department of Astrophysical Sciences, Princeton University, 4 Ivy Lane, Princeton, NJ 08540, USA}
\email{anatoly@astro.princeton.edu}

\begin{abstract}
    The diocotron instability is a nonaxisymmetric plasma instability that should occur generically in the differentially rotating equatorial plane of pulsar magnetospheres. We present a series of 3D particle-in-cell (PIC) simulations of the diocotron instability in aligned and oblique pulsars. The instability grows on timescales of the rotation period and develops a strong, stable $m=1$ mode, corresponding to a rotating, dipolar charge asymmetry in the equatorial disk. Stochastic fluctuations in the diocotron-mode amplitude and pattern speed drive cross-field diffusion that can rapidly transport charges through the closed zone toward the light cylinder. In the nonlinear stage, the $m=1$ mode produces electric field perturbations which can modulate the polar-cap potential drop and the emission-beam angle, with possible connections to pulsar variability such as nulling, periodic amplitude modulation, and drifting subpulses.
\end{abstract}


\section{Introduction}
Pulsars are rapidly rotating neutron stars which emit copious radiation across the electromagnetic spectrum from radio to gamma rays (see \citet{philippov_pulsar_2022} for a review). Since their discovery, extensive effort has been devoted to understanding the structure and dynamics of pulsar magnetospheres, where the broadband radiation originates. Much of the observed spectral, spatial, and temporal complexity is thought to arise from plasma processes in the magnetosphere, including plasma instabilities that drive time-dependent dynamics affecting the emission. 

One such instability is the diocotron instability, which derives its name from the Greek word for ``pursue.'' Diocotron modes are low-frequency, electrostatic drift waves in a nonneutral plasma \citep{Fine1988}. Diocotron is the electrostatic analog of the ideal Kelvin--Helmholtz instability \citep{chim_flux-driven_2016}, feeding on the free energy of differential rotation and evolving into a train of nonlinear, drifting vortices. 

The diocotron instability has been studied in detail in the context of laboratory plasmas, where it occurs in magnetized electron columns \citep[e.g.,][]{davidson_quasilinear_1985,Fine1988,fine_measurements_1989, driscoll_experiments_1990, davidson1990}. 
 In the astrophysical context, however, it has received comparatively little attention, despite the conditions for its onset arising naturally in pulsar magnetospheres. Ferraro's isorotation theorem requires plasma to corotate along a magnetic flux surface if the potential is constant along that surface \citep[e.g.,][]{Kulsrud2005}. Conversely, a potential drop implies noncorotation, corresponding in pulsar terminology to a ``gap'' where a parallel electric field remains unscreened. Plasma on field lines connected to the star through such gaps differentially rotates relative to the star, providing the conditions under which the diocotron instability develops. In this work, we investigate the development and nonlinear evolution of the diocotron instability in pulsar magnetospheres, finding that it efficiently diffuses charges through the magnetosphere and can significantly perturb plasma where radio emission originates. 

 This paper is structured as follows. Sections~\ref{sec:gaps} and \ref{sec:dioc_sims} review theoretical and numerical evidence for vacuum gaps and the diocotron instability in pulsars.
 In Section~\ref{sec:simulations}, we explain the setup of our 3D PIC simulations of the diocotron instability in pulsars. Section~\ref{sec:results} details the numerical results for the evolution of the instability. In Section~\ref{sec:discussion}, we discuss the effects of the instability on the magnetospheres of active pulsars and suggest possible observational consequences as well as future directions. In Appendices~\ref{appx:modeling_charge} and \ref{sec:diffusion_appendix}, respectively, we model the electric potential due to the diocotron modes and derive an estimate for the radial diffusion coefficient. Appendix~\ref{sec:pp_run} illustrates the structure of a diocotron-unstable magnetosphere with pair production close to the star.

 \subsection{Vacuum Gaps in Pulsar Models}\label{sec:gaps}

 \begin{figure}
     \centering
     \includegraphics[width=\linewidth]{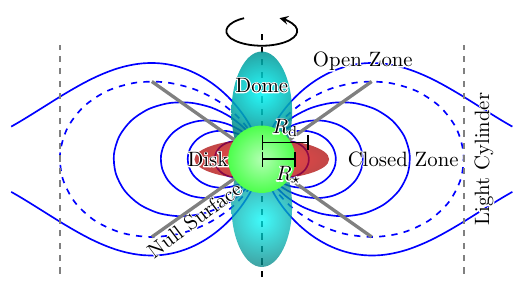}
     \caption{Schematic of a slice through an aligned, disk-dome pulsar magnetosphere. Blue lines represent magnetic flux surfaces. The dashed blue line touches the light cylinder (vertical gray dashed lines) and is the last closed field line, separating the closed zone from the open field lines. 
     Charges extracted from the surface of the pulsar, in green, form the equatorial disk, in red, and the domes above the poles, in cyan. The Goldreich--Julian charge density $\rho_{\rm GJ}$ changes sign at the null surfaces, represented by thick gray lines. 
     }
     \label{fig:cartoon}
 \end{figure}

 The early pulsar model of \citet{GJ1969} considered a corotating magnetosphere which is force-free, meaning that the electric field parallel to the magnetic field is everywhere screened. The charge density takes on the Goldreich--Julian (GJ) value $\rho_{\rm GJ}=-{\bf \Omega_\star\cdot B}/2\pi c$, where $\Omega_\star$ is the rotational frequency and the star subscript indicates quantities on the stellar surface at the pole. The GJ charge vanishes where ${\bf \Omega_\star\cdot B}=0$, referred to as the ``null surface.'' In an aligned pulsar with a dipole magnetic field, the null surface lies at an angle $\approx54.7^\circ$ from the rotation axis, illustrated by the thick gray lines in Figure~\ref{fig:cartoon}.
 
It was eventually realized that pulsar magnetospheres in the absence of pair formation may not be completely force-free and instead feature large vacuum gaps \citep{Holloway1973,Michel1979}.
The first self-consistent models of such a magnetosphere were calculated by \citet{KrausePolstorff1985a,KrausePolstorff1985b}. The corotating GJ charge density within the star produces a quadrupolar electric field that lifts electrons from the neutron star atmosphere around the magnetic poles and extracts ions from the equatorial regions. The extracted charges follow magnetic field lines until electrostatic equilibrium is reached. On field lines which pass through the null surface, the charges that are extracted at the base are of the wrong sign to satisfy the GJ charge density on the opposite side of the null surface. The magnetosphere therefore cannot attain a force-free state by extracting charges from the star \citep{Holloway1973}.  

 \citet{KrausePolstorff1985a,KrausePolstorff1985b} numerically calculated this charge extraction and obtained a solution that consists of an equatorial torus of ions called the ``disk'' and two blobs of electrons above the polar regions called ``domes.'' This configuration, called the disk-dome solution, features large gaps bounding the null surface, as illustrated in Figure~\ref{fig:cartoon}. Gaps are present in regions not filled by the disk or dome plasma.
 Because the outer region of the disk is threaded by magnetic field lines connected to the star through gaps, it is differentially rotating. 
 
\citet{petri_global_2002} later calculated a number of solutions with surface charge extraction and found that all should have superrotating regions in the disk. 
As demonstrated by these numerical results, the fact that the null surface intersects magnetic field lines means that a corotating, force-free GJ magnetosphere cannot be obtained by extracting charges from the stellar surface. Vacuum gaps and differential rotation must result. 

Crucially, large vacuum gaps around the null surface should form independently of the mechanism by which charges populate the magnetosphere. The reason is that, as theoretically predicted by \citet{Pilipp1974} and demonstrated numerically by \citet{Thacker1998} and 
\citet{smith_numerical_2001}, a corotating GJ magnetosphere of finite extent is unstable. A finite, force-free magnetosphere will collapse into a disk-dome-type solution, as arbitrarily distant quadrupolar charges are essential for the stability of the GJ solution. 

Pair production on the closed field lines can modify this collapse if the pairs can fill the field line with plasma and close the gap.
\citet{petri_global_2002} found that magnetic pair production becomes inefficient beyond a {\it breakdown radius} of $\sim 50 R_\star$ for $B_\star\sim10^{12}\,{\rm G}$, so the magnetosphere inside this radius is approximately force-free while the outer disk remains superrotating. This solution is demonstrated in Appendix~\ref{sec:pp_run}. Because the breakdown radius is much smaller than the light cylinder for $P\sim0.1$--$1\,{\rm s}$, gaps bound the null surface through most of the magnetosphere. The diocotron instability is therefore expected to occur generically on the closed field lines of nearly all nonmillisecond pulsars.

The linear properties of the diocotron instability in aligned rotators with surface charge extraction were analyzed by \citet{petri_diocotron_2002}. As expected, the superrotating equatorial disks are linearly unstable to the diocotron instability, with rapid growth rates on the order of the pulsar rotational period.  

\subsection{Simulations of the Diocotron Instability and Equilibrium Magnetospheres}\label{sec:dioc_sims}

 The diocotron instability was first observed in 3D PIC simulations by \citet{Spitkovsky2002} and subsequently in simulations of the disk-dome solution in \citet{philippov_ab_2014}. While PIC simulations are commonly used to study pulsar magnetospheres, the diocotron instability is not typically seen, for two reasons.

 First, 2D axisymmetric simulations are often used to study aligned pulsars. To minimize computational expense, works such as \citet{chen_electrodynamics_2014,chen_filling_2020,bransgrove_radio_2023, cruz_particle--cell_2024} employ 2D simulations. While many of these solutions feature gaps in the region of closed field lines \citep[e.g.,][]{chen_filling_2020}, the azimuthal diocotron instability naturally does not occur in these axisymmetric simulations.

Second, simulations typically fail to capture the scale hierarchy relevant to the diocotron instability. Computational expense forces the light cylinder to within a few stellar radii, so strong pair production at both the star and the light cylinder renders the magnetosphere nearly force-free \citep[e.g.,][]{philippov_ab_2014,philippov_ab_2015,bransgrove_radio_2023,Hakobyan2023_pulsar}, precluding diocotron growth.

 In realistic pulsars, however, the pair breakdown is separated from the light cylinder by orders of magnitude \citep{petri_global_2002}. Indeed, by imposing a radial cutoff on pair production, \citet{chen_filling_2020} obtained solutions for active pulsars with gaps bounding the null surface in the closed zone.

Unlike the canonical force-free pulsar magnetosphere, most real pulsars feature gaps around the null surface in the closed zone. 
Where the disk is threaded by field lines which pass through these gaps, the plasma loses corotation and is therefore susceptible to the diocotron instability. 
We denote this location as the {\it diocotron radius}, $R_{\rm d}$. Real pulsars satisfy the hierarchy $R_\star\ll R_{\rm d}\ll R_{\rm LC}$.
Directly simulating this scale separation is computationally expensive. In this work, we therefore model pulsars in which plasma is extracted from the stellar surface only, and the gaps in the closed zone extend down to the star. 
This is equivalent to setting the pair production threshold $\gamma_{\rm thr}\rightarrow\infty$, so that $R_{\rm d}\sim R_\star$. 
While this cannot capture active pulsars, it isolates the general evolution of the diocotron instability.

\begin{figure}[!htb]
    \centering
    \includegraphics[width=0.95\linewidth]{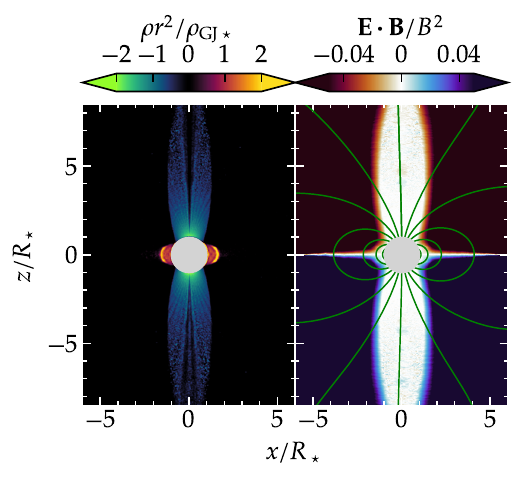}
    \caption{
Slices in the $x$--$y$ plane showing the charge density (left) and the parallel electric field (right) from a simulation of an aligned pulsar with surface charge extraction. (With strong cooling enabled, the electron dome does not reach steady state before the onset of the diocotron instability disrupts the disk. Therefore, this figure shows a run without cooling to illustrate the disk-dome equilibrium. The effect of cooling on the initial development of the dome does not affect the later evolution of the system.)
The snapshot is taken after the system has relaxed to an axisymmetric equilibrium but before the onset of the diocotron instability.
The charge density is normalized to the Goldreich--Julian density at the stellar surface and scaled by $(r/R_\star)^2$ for clarity.
The parallel electric field is normalized to the magnitude of the local magnetic field.
Sample magnetic field lines are overplotted in green.}
    \label{fig:fig1_sideview}
\end{figure}

\section{Simulation Setup}\label{sec:simulations}

We use the relativistic PIC code \textsc{ Tristan-MP v2} \citep{hayk_hakobyan_2023_7566725}. 
Our 3D Cartesian simulations have resolutions of $(n_x,n_y,n_z)=(804,804,1200)$ cells. The pulsar spin axis is oriented along $z$. The magnetization $\sigma\equiv B_\star^2/4\pi n_{\rm GJ\star} mc^2\approx 10^3$ in the GJ-density plasma on the stellar surface. The stellar radius is $R_\star=60\textrm{ cells}$, and $R_{\rm LC}/R_\star=7.17$. The GJ density on the stellar surface at the equator corresponds to 290 particles per cell, and, at this density, the skin depth is resolved with 3 cells. To avoid strong initial transients, we ramp up the spin over a period of $9 R_\star/c$. The outer boundaries employ absorbing conditions for the fields and outflow conditions for the particles. The star is modeled as a perfect conductor on which the magnetic field is forced to a dipole and the electric field to its corotating value. Particles are removed within the star. 
Within 10 cells of the star, or for $|{\bf E|/|B}|<0.95$, $c/\omega_c<2$, and $v_\perp/c<0.1$, particles are moved using a guiding-center-approximation (GCA) pusher, which allows us to under-resolve the gyroradius; otherwise, we use the Boris pusher. 

The plasma that populates the magnetosphere consists of electron--positron pairs that are extracted from the surface of the star. To ensure that surface charges are always available for extraction, we maintain a layer of plasma above the star, referred to as the ``atmosphere.'' Whenever the total number density in the atmosphere falls below $2.5n_{\rm GJ\star}$, we locally inject cold neutral plasma with particle weights proportional to the parallel electric field $|{\bf E\cdot \hat{B}}|$. To prevent charges from escaping when the parallel electric field is screened, we apply a small, inwards gravitational force in the layer. The atmosphere thickness is set to 1 cell in order to mimic a thin neutron star atmosphere.

We perform runs for an aligned as well as $5^\circ$ and $20^\circ$ oblique pulsars. In general, particles in the closed zone can move along the magnetic field lines, bouncing across the midplane in the electrostatically trapped disk. This bouncing can be resonant with certain diocotron modes. Deep within the light cylinder of real pulsars, the phase velocity of diocotron modes is $\ll c$, and energetic particles in the disk with velocities $v\sim c$ will not be resonant. Since our simulated pulsars rotate artificially quickly compared to normal pulsars in nature, we choose to avoid resonances by forcing the opposite hierarchy in which the bounce velocity is much smaller than the diocotron phase velocity. To achieve this, we strongly cool the positrons in the aligned rotators using the \citet{LL1975} radiation reaction formula. Due to their movement along curved field lines, the positrons rapidly lose their parallel momentum and do not resonate with diocotron modes.

In oblique rotators, the bounce motion is driven by the rotating magnetosphere rather than thermal velocity, so the strong cooling used for the aligned case would suppress this dynamics; we therefore apply only weak cooling there.

\section{Results}\label{sec:results}
\begin{figure*}
    \centering
    \includegraphics[width=0.9\textwidth]{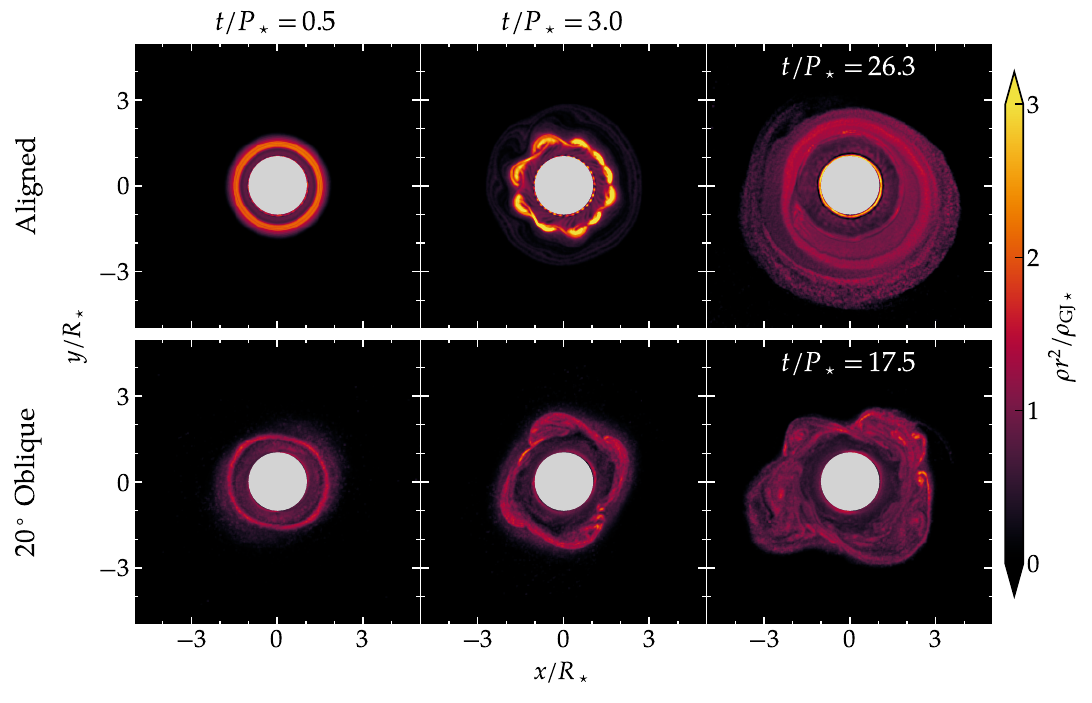}
    \caption{Slices through the magnetic equatorial plane of the charge density normalized to the Goldreich--Julian value at the stellar surface, scaled by a factor of $r^2$ for visual clarity. From left to right, panels show increasing time. The top and bottom rows correspond to the aligned and $20^\circ$ oblique runs, respectively. An animation of this figure is available at \href{https://youtu.be/nO2B5gv5Gww}{https://youtu.be/nO2B5gv5Gww}. It shows the temporal evolution of the charge density in slices through the equatorial and vertical planes for the aligned, $5^\circ$, and $20^\circ$ runs.}
    \label{fig:topview}

\end{figure*}

The basic disk-dome equilibrium, before the onset of the diocotron instability, is shown in Figure~\ref{fig:fig1_sideview}. As expected, we obtain two domes of electrons over the poles and an equatorial disk of positrons. The parallel electric field is screened on the surface and within the disk and dome. Some magnetic field lines which thread the disk connect to the star through vacuum gaps, causing the outer disk to rotate faster than the inner disk. This differential rotation renders the disk unstable to the diocotron instability \citep{davidson1990}. Figure~\ref{fig:topview} shows slices through the magnetic equator of charge density in the aligned and $20^\circ$ oblique runs, demonstrating the evolution of the diocotron instability. For all runs, the initial disk configuration at $t=0.5P_\star$ consists of corotating plasma up to $\sim1.5R_\star$, where a density spike is accompanied by a sharp gradient in the rotational frequency. 

\begin{figure*}
    \centering
    \includegraphics[width=0.9\textwidth]{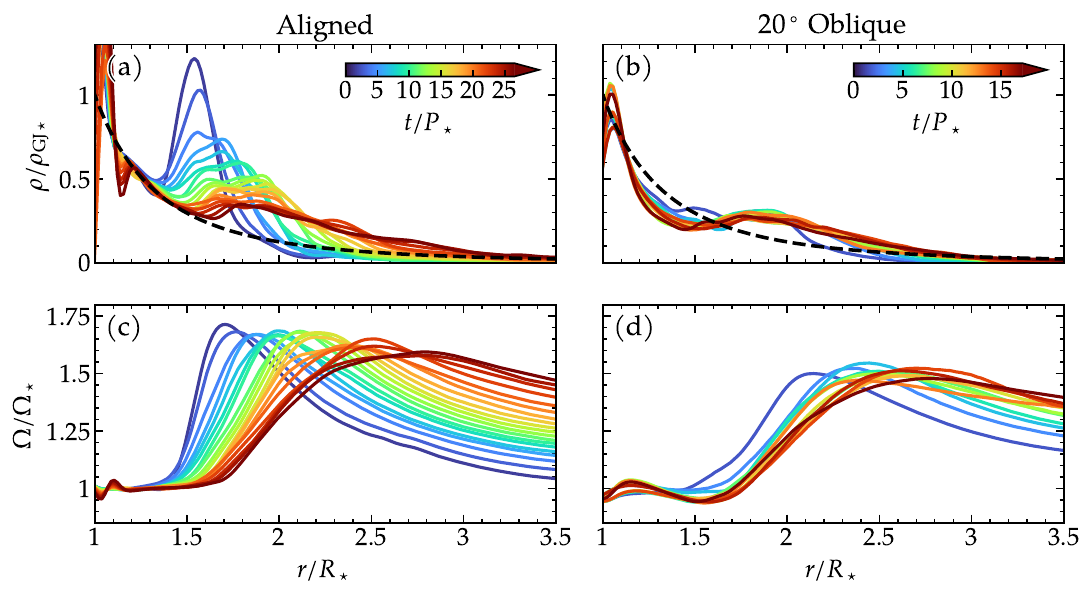}
    \caption{Radial profiles of azimuthally integrated quantities in the magnetic equatorial plane. Color indicates time as a fraction of pulsar period $P_\star$. The left column corresponds to the aligned rotator, and the right to the $20^\circ$ oblique rotator, both without pair production. (a), (b) The charge density normalized to the surface Goldreich--Julian density, with black dashed curves marking the local Goldreich--Julian density. (c), (d) The local rotational frequency as a fraction of the pulsar rotation frequency. }
    \label{fig:profile}
\end{figure*}

As expected, the diocotron instability grows on the timescale of the rotational period. The middle panels at $t=3P_\star$ show the instability in its linear stage. The aligned run is initially unstable to an $m=8$ mode, whereas the $20^\circ$ oblique run is immediately dominated by the $m=2$ mode. While the dominant mode in the linear stage depends on the details of the initial equilibrium, the evolution at late times does not. The disks become turbulent in the rotational plane as the vortices created by the diocotron instability turn over and merge. All disks are eventually dominated by low mode numbers ($m=1$ and 2). 
The azimuthal charge perturbations give rise to $E_\phi$ which radially rearranges the charge distribution, spreading it out and reducing the rotation frequency gradients. This radial charge transport is evident in Figure~\ref{fig:topview}, which shows the expansion of the disk with time.

The radial profiles in the equatorial plane of charge density and rotational frequency for the aligned and $20^\circ$ oblique runs are shown in Figure~\ref{fig:profile}. Dark blue curves correspond to the initial configuration. By comparing this initial state to the final states in red, it is clear that the diocotron instability pushes the charge density toward the GJ value and softens the rotational frequency gradient. Shallower rotational frequency gradients, corresponding to larger length scales, shift the fastest-growing diocotron mode to smaller $m$. By smoothing the rotation profile, charge transport due to the diocotron instability allows large-scale, small-$m$ modes to grow. Nonlinearly, this is achieved by the merging of vortices and results in a dominant $m=1$ mode at late times. This evolution agrees with the body of work on the diocotron instability in laboratory plasmas and will be discussed further in Section~\ref{sec:long_term}. 

Compared to the aligned and $5^\circ$ oblique runs, the $20^\circ$ oblique runs are initially unstable to lower-$m$ modes. Perturbations in the oblique rotators become nonlinear more quickly, with disks that appear more turbulent.

\subsection{Modeling Effects on the Magnetosphere}\label{sec:modeling_effects}
Here we develop a framework for quantifying the magnitude of the diocotron instability and its effect on the pulsar magnetosphere. We parameterize the charge distribution with two quantities: $|{Q}_m|$ is the total amount of charge participating in a diocotron mode with number $m$,  and $R_{\rm d}$ is the average radius of charges in a given diocotron mode and can shift as the instability redistributes charge. $R_{\rm d}$ must enclose some field lines which pass through gaps near the null surface. We obtain ${Q}_m$ by integrating the charge density against spherical harmonics $Y_{\ell m}(\theta,\phi)$. Since the diocotron pattern has one poloidal node and $m$ azimuthal nodes, we set $\ell=m$ and use the shorthand $Y_{\ell m}=Y_m$. Then,
\begin{equation}\label{eq:Qm}
    {Q}_m\equiv\int dV \,\rho(r,\theta,\phi)\, Y_m^\star(\theta,\phi).
\end{equation}
This quantity can be measured in our simulations by performing the integral over the charge distribution. 

\begin{figure*}
    \centering
    \includegraphics[width=0.9\textwidth]{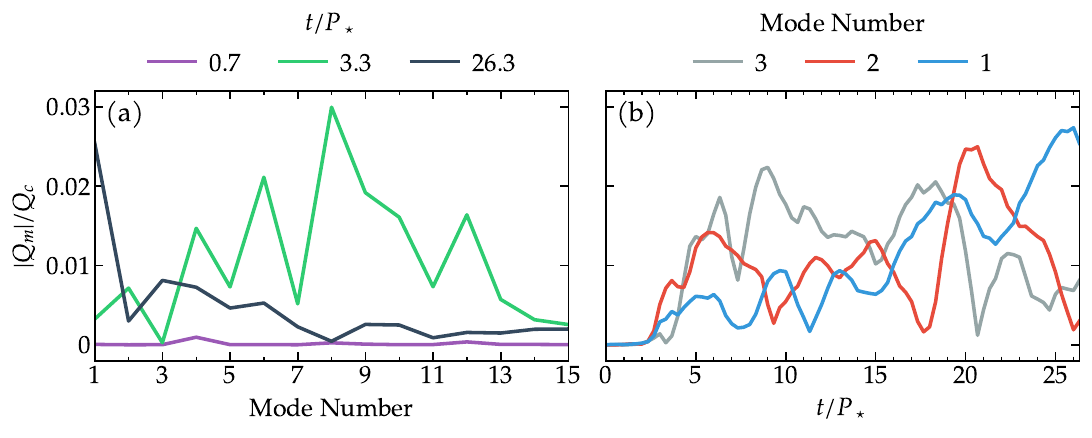}
    \caption{The disk charge contained in angular mode $m$ (see Section~\ref{sec:modeling_effects}) normalized to the central point charge $Q_c$, in the aligned simulation. (a) As a function of mode number $m$ for times $t/P_\star=0.7$, 3.3, and 26.3. The initial equilibrium (purple, $t=0.7P_\star$) is axisymmetric with little power in any mode. The $m=8$ mode dominates in the linear stage (green, $t=3.3P_\star$). At late times (dark gray), the nonlinear evolution results in a dominant $m=1$ mode and little power in higher-$m$ modes.
    (b) As a function of time for modes $m=1$, 2, and 3. All modes start to grow at $t\approx2P_\star$. As nonlinear vortices develop and merge, energy is exchanged between the $m=1$, 2, and 3 modes, with the $m=1$ mode dominating at late times.   }
    \label{fig:multipoles}
\end{figure*}

We approximate the radial dependence of the nonaxisymmetric charge distribution as a Dirac delta function, thereby avoiding the need to model a complicated radial structure. 
Using these measured quantities, we write the charge distribution as a sum of spherical harmonics with radial dependence given by a Dirac delta at $R_{\rm d}$: 
\begin{equation}\label{eq:rho_model}
    \rho(r,\theta,\phi)=\sum_m{{Q}_m}\,Y_{m}(\theta,\phi)\frac{\delta(r-R_{\rm d})}{r^2}.
\end{equation}
In Figure~\ref{fig:multipoles}, we measure $|Q_m|$ in the aligned run as a function of mode number and time, normalized to the central point charge $Q_c\equiv (2\pi/3)\rho_{\rm GJ\star}R_\star^3$ \citep{Michel1991_book}. In panel~(a), we see that the $m=8$ mode dominates in the linear stage around $t=3.3 P_\star$. At late times, vortices have merged so that the charge distribution is dominated by $m=1$ and $m=3$ modes. Panel~(b) illustrates how the low-$m$ modes grow rapidly after a few $P_\star$. As charge vortices merge, the $m=1$ grows more slowly than $m=2$ and 3. The mode amplitudes vary nonmonotonically in time, but there is a clear trend toward $m=1$ dominance. In all cases, $|Q_m|$ for the dominant mode is a few percent of $Q_c$. 
The mode amplitude should remain similar even if pair production fills the magnetosphere out to a finite radius (see Appendix~\ref{sec:pp_run}), since the solution is expected to remain approximately self-similar: The volume of the torus containing the charge fluctuations scales as $r^3$, while the GJ charge density falls as $r^{-3}$. 

Figure~\ref{fig:potential_vis}(b) shows the pattern rotational frequencies for the $m=1$, 2, and 3 modes in the aligned pulsar. There are significant fluctuations in the linear stage, but the modes approach a frequency of $\omega_m/m\sim1.2\,\Omega_\star-1.3\,\Omega_\star$.  

\section{Discussion}\label{sec:discussion}

\subsection{Evolution Toward the $m=1$ Mode}\label{sec:long_term}
Due to the computational expense, we ran our simulations for only tens of rotational periods, corresponding to real timescales of seconds to a minute. Real pulsars face no such limitation, so it is important to establish the theoretical late-time behavior of the diocotron instability. We find that the magnetosphere should develop a robust $m=1$ mode, which can be understood in two ways.

First, the secular evolution of the magnetosphere naturally drives the system toward conditions favorable for the $m=1$ instability. As shown in Figures~\ref{fig:profile}(c) and (d), the diocotron instability progressively flattens the differential rotation profile. Since the azimuthal wavenumber of the fastest-growing mode scales with the steepness of the shear, a flatter rotation profile shifts the peak growth rate to smaller $m$ \citep{petri_diocotron_2002,petri_diocotron_2007}.

Second, the $m=1$ mode results from the merging of charge-overdense vortices. Because the diocotron instability itself relies on the effective mutual attraction of charges in the disk, overdensities attract and merge, shifting power to lower-$m$ modes.
Our simulations indeed show rapidly merging vortices that shift the dominant mode to lower $m$. 
Vortex merging and rotation-profile flattening both favor $m=1$, and, once established, the mode persists. Such stability is well documented in laboratory experiments and the associated theory, which have long recognized the $m=1$ mode as a stable result of the nonlinear diocotron evolution \citep[e.g.,][]{Briggs1970,fine_measurements_1989,davidson_equilibrium_1991,oneil_stability_1992,kawai_relaxation_2006}.  

Indeed, the $m=1$ mode {\it must} persist in pulsars unless the closed zone of the magnetosphere reaches an entirely force-free state, since any closed zone with gaps will exhibit differential rotation and remain diocotron-unstable. Theoretical and numerical results indicate that such a force-free magnetosphere of finite extent is itself unstable and will always form gaps around the null surface in the absence of efficient local pair production \citep{Pilipp1974,Thacker1998,smith_numerical_2001}. When pairs cannot be produced locally around the null surface, vacuum gaps open, sustaining a diocotron-unstable disk that tends toward the $m=1$ state. 

\subsection{Radial Charge Transport}
\begin{figure}[tb]
    \centering
    \includegraphics[width=\linewidth]{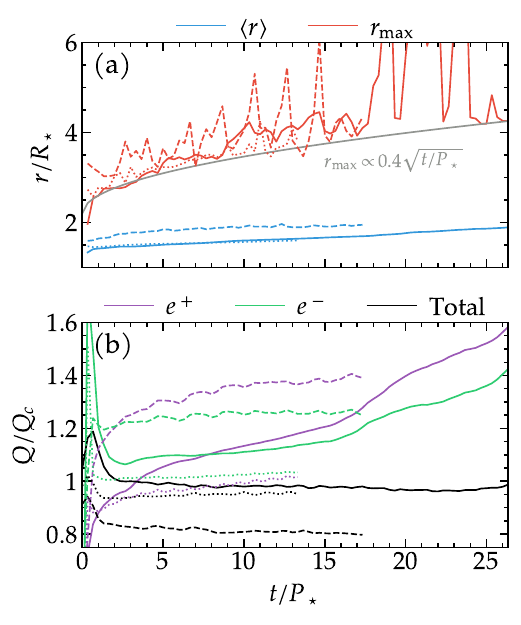}
    \caption{Diagnostics of diffusive radial transport. Solid, dotted, and dashed curves correspond to the aligned, $5^\circ$, and $20^\circ$ oblique simulations, respectively. (a) In blue, the average radius of positrons in the disk with $r>1.1R_\star$, denoted $\langle r\rangle$. In red, the maximum radius at which the positron density exceeds $0.03\rho_{\rm GJ\star}$, denoted $r_{\rm max}$. 
    (b) Purple and green show the total number of positrons and electrons, respectively, with $r>1.1R_\star$, to exclude charges on the stellar surface. This demonstrates the steady increase in the number of particles in the disk and dome. The black curve is the total electric charge in the system. It is roughly constant, meaning that there is no net current in the magnetosphere. All quantities are normalized to the central-point-charge magnitude $Q_c$. }
    \label{fig:rad_transport}
\end{figure}
A longstanding question in pulsar magnetosphere theory is whether vacuum gaps persist in the outer closed zone \citep[e.g.,][]{Holloway1973}. In weak pulsars without pair production near the light cylinder, charges enter the magnetosphere either by extraction from the stellar surface or by magnetic pair production in the inner magnetosphere at $r\lesssim 50R_\star$ \citep{petri_global_2002}. Because the null surface intersects the field lines along which charges flow, these sources cannot supply the outer magnetosphere with the correct sign of charge.  

This raises the question of whether cross-field transport can fill the outer magnetosphere. Nonaxisymmetric diocotron modes generate an azimuthal electric field $E_\phi$ that produces radial ${\bf E}\times{\bf B}$ drift.

In the frame corotating with the $m=1$ pattern, drifting disk charges follow closed equipotential surfaces, resulting in closed orbits with zero net radial displacement. However, stochastic fluctuations in the mode strength and pattern speed perturb the equipotential surfaces, preventing orbits from closing and leading to quasilinear radial diffusion. The dominant effect in our simulations is from fluctuations in the $m=1$ mode strength (see Figure~\ref{fig:multipoles}(b)).

In Appendix~\ref{sec:diffusion_appendix}, we obtain the corresponding diffusion coefficient by integrating the radial velocity autocorrelation function: $D = \int_0^\infty d\tau\, \langle V_r(t)V_r(t+\tau)\rangle$. The result is
\begin{align}
D_r 
&\approx 
\frac{\bar{V}_{r0}^2\tau_{c,Q_1}}{2}\frac{\langle \delta Q_1^2/\bar{Q}_1^2\rangle}{1+\tau_{c,Q_1}^2\Delta\bar{\omega}_1^2}\notag \\
&=
\frac{4\pi}{27}
{R_{\rm d}^2\Omega_\star^2}{\tau_{c,Q_1}}
\frac{\langle \delta Q_1^2/{Q}_c^2\rangle}{1+\tau_{c,Q_1}^2\Delta\bar{\omega}_1^2} ,
\label{eq:diff_coeff}
\end{align}
where $\bar{V}_{r0}$ is the characteristic velocity of radial kicks from the diocotron mode, $\Delta\bar{\omega}_1=\Omega(r)-\bar{\omega}_1$ is the difference between the local rotational frequency and the average $m=1$ mode pattern frequency, 
$\langle \delta Q_1^2\rangle$ is the variance of the mode charge amplitude, 
and $\tau_{c,Q_1}$ its correlation time.

From the aligned simulation (for $t>14.1P_\star$), we measure
\begin{equation*}
\left\langle \delta Q_1^2/Q_c^2 \right\rangle 
\simeq 4\times10^{-5},
\qquad
\tau_{c,Q_1}\sim P_\star .
\end{equation*}
Rotational frequencies in the disk have variations of order $10\%$, so $\Delta\bar{\omega}_1/\Omega_\star\sim0.1$ and $\Delta\bar{\omega}_1\tau_{c,Q_1}\lesssim1$. Evaluation of Equation~(\ref{eq:diff_coeff}) with the reasonable approximation $(\tau_{c,Q_1}\Delta\bar{\omega}_1)^2\ll 1$ yields
\begin{align}\label{eq:diff_co_numerical}
D_r 
&\approx 
\frac{4\pi}{27}R_{\rm d}^2\Omega_\star^2\tau_{c,Q_1}\left\langle\frac{\delta Q_1^2}{{Q}_c^2}\right\rangle \\
&\sim
 10^{-3}
\frac{R_{\rm d}^2}{P_\star}\notag
 \\
&\sim 
2
\frac{R_\star^2}{P_\star}
\left(\frac{R_{\rm d}}{50R_\star}\right)^2 .\notag
\end{align}
The dipolar electric field exterior to the $m=1$ mode is proportional to $R_{\rm d}^3$, while the background magnetic field falls with $r^{-3}$. Therefore, the diffusion efficiency increases with $R_{\rm d}$ in Equation~(\ref{eq:diff_co_numerical}).

\paragraph{ Comparison with simulations.}
In Figure~\ref{fig:rad_transport}(a), we plot the average and maximum radii of positrons in the disk. Both increase with time for the aligned, $5^\circ$, and $20^\circ$ rotators. In particular, the maximum radius $r_{\rm max}$ grows with $\sqrt{t}$, consistent with diffusion. As disk charges are transported outward, additional charges are extracted from the star to ensure the corotating inner disk is filled with GJ-density plasma.
% The corresponding increase of total charge in the disk and the dome is demonstrated by the purple and green curves 
In Figure~\ref{fig:rad_transport}(b), we plot the total charge of positrons and electrons in the disk and domes in purple and green, respectively, with the total system charge in black. The total system charge remains constant, while the increasing number of positrons and electrons indicates the progressive expansion of the disk and domes. The spreading of the radial charge profile is also evident in Figures~\ref{fig:profile}(a) and (b). While the transport properties for oblique rotators seem similar to the aligned case, we did not run the oblique simulations for long enough to draw conclusions about how the diffusive transport depends on pulsar obliquity.

{}
{}
We use Fick's law with the azimuthally averaged charge-density profiles in Figure~\ref{fig:profile}(a) to obtain a best-fit diffusion coefficient $D_r=\langle-J_r/\partial_r\rho\rangle_t$, averaged over the nonlinear stage $t>14.1P_\star$ in the aligned simulation. The typical value is $D_r\sim2\times 10^{-2}R_\star^2/P_\star$.  
With $R_{\rm d}\approx 2R_\star$, the simulation diffusion coefficient exceeds the estimate in Equation~(\ref{eq:diff_co_numerical}) by a factor of a few. This discrepancy can be accounted for by diffusion due to the higher-$m$ modes (e.g., $m=2$, 3, and 4) that still have significant amplitude at the end of the simulation (see Figure~\ref{fig:multipoles}).

Unlike earlier quasilinear and 2D numerical studies \citep{petri_cross-field_2003, petri_non-linear_2009}, which required prescribed particle injection, our 3D simulations demonstrate self-consistently that nonlinear diocotron fluctuations drive efficient diffusive transport through the closed magnetosphere.

\paragraph{ Filling the closed magnetosphere.} 
Diffusion acts to smooth the outwardly decreasing GJ density ($\rho_{\rm GJ}\propto r^{-3}$), producing a net outward radial flux of particles. The diffusive current density is
\begin{equation}
J_r \sim -D_r \frac{\partial\rho}{\partial r}
\sim \frac{3D_r\rho_{\rm GJ}}{r}.
\end{equation}
Integrating the current density within the null surface ($\cos\theta\in[-1/\sqrt{3},1/\sqrt{3}]$)  at $\sim R_{\rm d}$ yields the total radial current:
\begin{equation}
I_{\rm diff}
\sim 4\sqrt{3}\,
\frac{Q_c D_r}{R_{\rm d}^2}.
\label{eq:I_diff}
\end{equation}

The charge required to fill the region between $R_{\rm d}$ and the light cylinder $R_{\rm LC}$ with $\rho_{\rm GJ}$ is
\begin{equation}
Q_{\rm fill}
\sim 
\frac{4}{\sqrt{3}} Q_c 
\ln\!\left(\frac{R_{\rm LC}}{R_{\rm d}}\right).
\label{eq:charge_left}
\end{equation}

The corresponding filling time is
\begin{align}
t_{\rm fill}
&\sim 
\frac{Q_{\rm fill}}{I_{\rm diff}} \\
&\sim 
\frac{1}{3}
\ln\!\left(\frac{R_{\rm LC}}{R_{\rm d}}\right)
\frac{R_{\rm d}^2}{D_r}.
\label{eq:tfill}
\end{align}
For $P_\star\sim1\,{\rm s}$, this yields $t_{\rm fill}$ of order an hour, consistent with extrapolating the growth of $r_{\rm max}$ in Figure~\ref{fig:rad_transport}(a).
{}

\subsection{$m=1$ Mode Morphology and Amplitude}\label{sec:mode_morphology}

In the absence of a saturation mechanism, we expect the radial amplitude of the $m=1$ mode to grow with time until it reaches the light cylinder. Charges extracted from the outer edge of the corotating region move 
to join the $m=1$ charge overdensity, while the outer edge of the corotation region diffuses slowly outwards. 
The radius of the disk and its maximum extent $r_{\rm max}$ grow rapidly, while the minimum extent of the nonaxisymmetric disk $r_{\rm min}$, corresponding to the underdensity of the diocotron mode, remains close to the star.
Figures~\ref{fig:profile}(a) and (c) demonstrate this process by showing how the maximum radius of the charge distribution increases more rapidly than the outer radius of the corotating plasma. Indeed, Figure~\ref{fig:topview} illustrates how the $m=1$ mode in the aligned case develops an extremely nonlinear shape similar to an offset cylinder.

How $R_{\rm d}$ and $|Q_1|$ change as the disk grows depends on the disk geometry.
For a cylindrical disk with $r^{-3}$, $|Q_1|\propto 1-r_{\rm min}/r_{\rm max}$ and the average radius $R_{\rm d}\approx \langle r\rangle\sim r_{\rm min}\ln(r_{\rm max}/r_{\rm min})$.  
For spherical geometry, $|Q_1|\propto\ln(r_{\rm max}/r_{\rm min})$, and $\langle r\rangle\propto (r_{\rm max}-r_{\rm min})/\ln(r_{\rm max}/r_{\rm min})$. These two scalings converge at small $r_{\rm max}/r_{\rm min}$, so our simulations cannot distinguish them. However, the disk does remain geometrically flat throughout, consistent with the cylindrical case.

\subsection{Diocotron Modes Near the Light Cylinder}
As diocotron modes diffusively transport charge outward through the magnetosphere, the outer edge of the superrotating, unstable region of the disk migrates toward the light cylinder.
Because plasma at the outer edge of the disk superrotates relative to the star, the radius at which the outer edge reaches $v_\phi \sim c$ lies inside the fiducial light-cylinder radius, $R_{\rm LC}=c/\Omega_\star$. Particles that are kicked by diocotron perturbations into the region where $v_\phi \gtrsim c$ (equivalently, where $E > B$) are ejected from the disk into the pulsar wind. This effectively sets a maximum radial extent for the disk.

The diocotron instability relies on rotational frequency gradients that allow charges to ``catch up'' with charge-density perturbations. Imposing a relativistic speed limit suppresses this mechanism. \citet{petri_relativistic_2007} analyzed the linear properties of the diocotron instability in the relativistic rotation regime and found increasing suppression as the maximum rotational velocity approaches the speed of light. In particular, some modes---preferentially at low $m$---can be completely stabilized when $v_\phi \lesssim c$ at the outer edge. 
If, however, $r_{\rm min}$ cannot efficiently diffuse toward the light cylinder and remains close to the star, like in our simulations, a differential rotation profile can persist in the disk, allowing the mode to exist near the light cylinder in a saturated state. A preexisting large-amplitude mode may therefore be resilient upon reaching the light cylinder. Analysis of the diocotron mode evolution near the light cylinder will be the subject of future work.

\subsection{Effects on Open Field Lines }\label{sec:pc_effects}
\begin{figure*}
    \centering
    \includegraphics[width=0.9\linewidth]{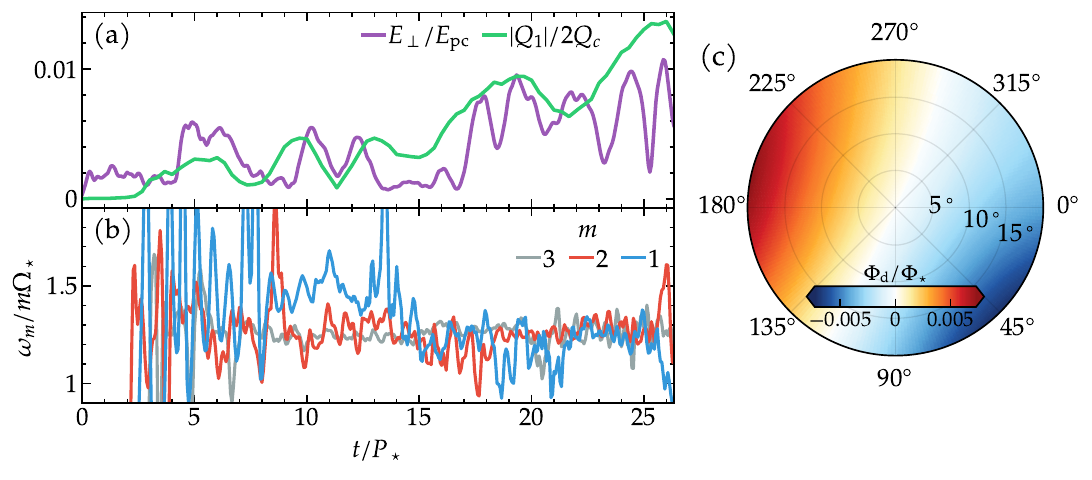}
    \caption{Diagnostics of the diocotron potential perturbation effects on the aligned pulsar. (a) In purple, time series of the perpendicular electric field $E_\perp=\sqrt{E_x^2+E_y^2}$ averaged on the polar axis from $z=1.1$ to $1.8R_\star$, normalized to the polar-cap electric field $E_{\rm pc}\equiv \rho_{\rm GJ\star}R_{\rm pc}$. In green, the normalized $m=1$ diocotron mode strength $|Q_1|/Q_c$, scaled down by a factor of 2 for clarity. (b) Time series of the pattern frequencies $\omega_m$ of the $m=1$, 2, and 3 modes, normalized to the stellar rotation frequency $\Omega_\star$. The pattern frequency is the time derivative of the complex phase of the multipole moment. For readability, the data are smoothed with a Gaussian filter with a standard deviation of $0.07\,P_\star$. (c) Polar diagram of the electric potential due to the diocotron charge perturbations from the $m=1$--5 modes, evaluated at the pulsar surface around the north pole and normalized to the full pulsar potential drop $\Phi_\star\sim2\pi\rho_{\rm GJ\star}R_\star^2$. Note that the polar-cap potential drop $\Phi_{\rm pc}$ is smaller than $\Phi_\star$ by a factor of $R_\star/R_{\rm LC}\ll 1$. The azimuthal coordinate is the longitude, and the radial coordinate is the colatitude from the pole. To illustrate, the data are evaluated at the final simulation snapshot (with the corresponding equatorial charge density in Figure~\ref{fig:topview}), but the pattern changes with time.   }
    \label{fig:potential_vis}
\end{figure*}

We now explore if the time-dependent nonaxisymmetric diocotron modes can meaningfully perturb the open-field-line bundle from which the radio emission originates.  

Consider the $m=1$ mode, since it is the only mode with nonvanishing electric field above the pole.
In Appendix~\ref{appx:modeling_charge}, we use the multipole expansion of the charge distribution in Equation~(\ref{eq:rho_model}) to estimate the diocotron potential perturbation due to the $m=1$ mode as 
\begin{equation}\label{eq:dioc_potential}
    \delta\Phi_{\rm d,1}=\sqrt{\frac{8\pi}{3}}\frac{|Q_1|}{R_{\rm d}}\sin\theta\,\cos\varphi_1  \begin{cases}
        \left(\frac{r}{R_{\rm d}}\right) & r\lesssim R_{\rm d}\\
         \left(\frac{r}{R_{\rm d}}\right)^{-2} & r\gtrsim R_{\rm d}
    \end{cases}
\end{equation}
The potential perturbation vanishes on the pole ($\theta=0$), so we evaluate it at the edge of the open-field-line bundle at radius $r$, where $\theta\approx\sqrt{r/R_{\rm LC}}$.
The magnitude of the perturbation to the potential drop between $R_\star$ and $r\gg R_\star$, compared to the fiducial polar-cap potential drop $\Phi_{\rm pc}\sim\rho_{\rm GJ\star} R_{\rm pc}^2$, is
\begin{equation}\label{eq:delta_phi_phiPC}
\begin{split}
\frac{\delta\Phi_{\rm d,1}}{\Phi_{\rm pc}}&\sim\frac{|Q_1|}{Q_c}\left(\frac{R_{\rm LC}}{R_{\rm d}}\right)^{1/2}\left(\frac{r}{R_{\rm d}}\right)^{3/2}\\
&\sim 0.1\left(\frac{|Q_1|}{0.02Q_c}\right)\left(\frac{R_{\rm LC}}{25R_{\rm d}}\right)^{1/2}\left(\frac{r}{R_{\rm d}}\right)^{3/2}
\end{split}\end{equation}
This quantity is sensitive to $R_{\rm d}$. For a cylindrical disk spanning the breakdown radius ($\sim 50R_\star$) to $R_{\rm LC}$, we have $R_{\rm d}\sim 200R_\star$. This also results in $R_{\rm d}$ close to typical emission altitudes \citep[e.g.,][]{Kijak2003}. With emission altitudes $r_{\rm em}\sim 0.25R_{\rm d}$ and for $|Q_1|/Q_c\sim 0.01$, this results in perturbations of magnitude $\sim0.01$ to the potential drop between the stellar surface and the emission region. 
If the disk is inclined to the emission beam by angle $\theta_{\rm disk}$, however, then the potential perturbation in Equation~(\ref{eq:delta_phi_phiPC}) is enhanced by a factor of $\theta_{\rm disk}/\theta_{\rm pc}$. Given typical polar-cap opening angles of $\sim1^\circ$, a disk slightly inclined by $\sim10^\circ$ could exert perturbations to acceleration in the emission region of order $0.1$. Such a precessing disk may arise in an oblique pulsar where the disk inclination is not rigidly coupled to the field lines. The inclined-disk scenario also enables diocotron modulation of the potential drop at the emission-beam core.

The $m=1$ mode also perturbs the electric field components $E_\theta\sim E_\phi$ on the open field lines. Taking the gradient of Equation~(\ref{eq:dioc_potential}) and comparing it to the characteristic electric field at the edge of the open-field-line emission region $E_{\theta,{\rm em}}\sim \rho_{\rm GJ\star}R_\star(R_\star/r)^2(r/R_{\rm LC})^{1/2}$ for an aligned pulsar gives
\begin{align}\label{eq:delta_Etheta}
   \frac{\delta E_{\phi}}{E_{\theta,{\rm em}}}\sim \frac{\delta E_{\theta}}{E_{\theta,{\rm em}}}\sim \frac{|Q_1|}{Q_c}\left(\frac{r}{R_{\rm d}}\right)^{3/2}\left({\frac{R_{\rm LC}}{R_{\rm d}}}\right)^{1/2}.
\end{align}

In a cylindrical disk with $R_{\rm d}$ near the emission height as before and $|Q_1|\sim0.01\,Q_c$, we obtain $\delta E_\phi\sim\delta E_\theta\sim E_{\theta,{\rm pc}}$. These perturbations could shift the emission beam by an angle
\begin{equation}\label{eq:deflection_angle}
\begin{split}
    \frac{\delta\theta}{\theta_{\rm em}}&\sim\left(\frac{\delta E_\theta c}{Br}\frac{2\pi}{\Delta\bar{\omega}_1}\right)\left(\frac{R_{\rm LC}}{r}\right)^{1/2}\\
    &\sim \left(\frac{|Q_1|}{0.02Q_c}\right)\left(\frac{\Delta{\bar{\omega}}_1}{0.1\Omega_\star}\right)^{-1}\\
    &\quad\times\left(\frac{r}{R_{\rm d}}\right)^{3/2}\left(\frac{200R_\star}{R_{\rm d}}\right)^{1/2} P_1^{1/2},
\end{split}
\end{equation}
where $r$ corresponds to the emission height, $\Delta\bar{\omega}_1\equiv\omega_1-\Omega_\star$ is the frequency of the $m=1$ pattern as seen in the stellar corotating frame, and $\theta_{\rm em}=(r/R_{\rm LC})^{1/2}$ is the local opening angle of the radio-emitting open-field-line bundle.
Therefore, a persistent $m=1$ diocotron mode could lead to significant (order-unity) deflections in the polar-cap field lines around the emission region.

Because charges in the highly magnetized magnetosphere are restricted to move along field lines, the dome charges are unable to effectively screen the electric field perturbation from reaching the polar cap. A simple upper-bound estimate derived in Appendix~\ref{sec:screening} demonstrates that this screening can reduce the field on the pole by at most a factor of order unity.

Figure~\ref{fig:potential_vis}(a) demonstrates how the $m=1$ diocotron mode modulates the electric field above the polar cap in our aligned simulation. The purple curve in panel (a) shows the perpendicular electric field $E_\perp=\sqrt{E_x^2+E_y^2}$ (corresponding to $E_{\theta,{\rm pc}}$) evaluated over the polar cap as a function of time. We also show in green $|Q_1|/Q_c$, as in Figure~\ref{fig:multipoles}(b), scaled down by a factor of 2 for visual clarity. There are two things to note. First, the magnitude of $E_\perp/E_{\rm pc}$ is consistent with the estimate from Equation~(\ref{eq:delta_Etheta}) and with the lack of screening. Second, the fluctuations in $E_\perp$ are clearly correlated with the strength of the $m=1$ diocotron mode. This is strong evidence that polar-cap electric field fluctuations are modulated by the $m=1$ diocotron mode. 

Figure~\ref{fig:potential_vis}(c) shows the potential perturbation due to the $m=1$--5 diocotron modes, evaluated over the pole at the final time step of the aligned simulation. While the specific pattern varies with the amplitudes and relative phases of the modes, this illustrates general features of the potential perturbation, such as its amplitude of $\lesssim0.01\Phi_\star$ (where $\Phi_\star\sim2\pi\rho_{\rm GJ\star}R_\star^2$) and the dominance of the dipolar $m=1$ mode.
\subsection{Pulsar Phenomenology}
Pulsar radio emission exhibits a remarkable diversity of poorly understood variable phenomena. Here we speculate on how some of these observations may be connected to the diocotron instability. Diocotron modes can influence radio emission through three distinct channels: by deflecting the beam direction via the poloidal and azimuthal electric field, by modulating acceleration and pair creation above the polar cap through nonaxisymmetric radial electric fields, and by driving charge diffusion through the closed zone, leading to large-scale changes in magnetospheric structure.

Nulling, in which pulsar radio emission ceases for some period of time, has been observed in over 200 pulsars \citep{Backer1970,wang_pulsar_2007,Wen2016}. It seems to come in two distinct varieties. Short nulls, lasting one to a few rotation periods, may arise when the emission beam swings away from the observer's line of sight, although disruptions to the emission process are also a viable mechanism. Long nulls, lasting hundreds to thousands of periods or more, are associated with large-scale magnetospheric changes and a measurable decrease in spin-down rate, suggesting a disruption of the pair supply rather than a simple geometric effect \citep[e.g.,][]{Kramer2006,Herfindal2009,Gajjar2014,Basu2017_nulling}. 

The deflection of the emission beam by the $m=1$ diocotron mode, as discussed in Section~\ref{sec:pc_effects}, can naturally account for the short-null variety. Roughly $35\%$ of nulling pulsars null periodically, with recurrence timescales of tens of periods, and, compared to sporadic nullers, these periodic nullers tend to have longer individual nulls \citep{Basu2017_nulling, Basu2020,Anumarlapudi2023}.
Both features are consistent with the diocotron picture: A smaller corotating pattern frequency $\Delta\bar{\omega}_1$ produces stronger beam deflection and longer recurrence times. Deflection of the entire beam by the diocotron mode also explains why periodic nulling is observed in both the core and cone components of the pulse profile, a result that is difficult to accommodate in carousel models \citep{Basu2017_nulling}. The stochastic character of diocotron modes can further account for the observed distribution of nulling periods, unlike deterministic mechanisms such as free precession \citep{Basu2017_nulling}. Long nulls, by contrast, are more plausibly connected to the slower, diffusive timescale on which diocotron-induced charge transport modifies the charge structure of the closed magnetosphere.

Closely related to nulling is periodic amplitude modulation (PAM), in which the pulse intensity varies periodically on timescales of order $10P_\star$, sometimes with pulse-phase-dependent structure \citep{Basu2020,basu_characterizing_2025}. PAM is consistent with what one would expect from a slightly superrotating diocotron mode that perturbs the accelerating potential above the polar cap and deflects the beam direction. The global nature of the diocotron perturbation naturally explains why PAM is observed simultaneously across both core and cone emission components, as well as in the interpulse. These properties are again difficult to reproduce in carousel models. 
For example, the central component of the Vela pulsar pulses displays modulations in intensity and longitude with period $\approx 9P_\star$ \citep{Wen2020}. 
The longitudinal drift of a few degrees and the periodicity of $9P_\star$ are both plausibly consistent with beam deflection (see Equation~(\ref{eq:deflection_angle})) due to a diocotron mode rotating with $\omega_1\approx 1.1\Omega_\star$.

More speculatively, drifting subpulses, present in $\sim60\%$ of pulsars \citep{Song2023}, may also reflect nonaxisymmetric diocotron perturbations, though we leave detailed treatment to future work.

Pulsar emission variability is exceptionally complex, and considerable further work is needed to connect diocotron modes to observations across the full pulsar population. Nevertheless, the diocotron instability has several properties that make it a promising unifying framework. It is time dependent, nonaxisymmetric, and capable of perturbing both the local emission region and the global magnetospheric structure. 
It is inherently nonlinear and stochastic, consistent with the complex behavior that characterizes pulsar variability \citep{Cordes2013}.

\section{Conclusion}

We performed a series of 3D PIC simulations of the diocotron instability in aligned and oblique pulsar magnetospheres. Consistent with theoretical and experimental expectations, diocotron modes grow quickly in regions of the equatorial disk connected to the star by field lines that cross vacuum gaps. In the nonlinear stage, vortices merge until the disk is dominated by a long-lived $m=1$ mode carrying a charge equivalent to a few percent of $Q_c\equiv (2\pi/3)\rho_{\rm GJ\star}R_\star^3$. While radial particle orbits in the diocotron potential are nearly closed, stochastic fluctuations in the mode strength and pattern speed lead to radial diffusion, which can efficiently transport plasma across field lines within the closed magnetosphere. From our simulations, we derive a diffusion coefficient indicating that typical pulsars can transport plasmas to the light cylinder on timescales of order hours. 

The persistent $m=1$ mode produces electric field perturbations that should significantly perturb the closed-field-line bundle near the emission region. Especially with a slightly inclined disk, the diocotron mode can induce fluctuations in the potential drop between the polar cap and emission region of order 0.01--0.1
times the fiducial polar-cap potential drop, possibly affecting particle acceleration.  
In addition, the electric field associated with the $m=1$ mode perturbs the azimuthal and polar field-line velocities near the emission region, tilting the emission beam. 
By simultaneously modulating beam orientation and polar-cap acceleration, the diocotron instability provides a compelling physically motivated mechanism for low-frequency pulsar emission variability, including nulling and periodic amplitude modulation.

Although we have focused on inactive disk-dome pulsars, the instability should arise generically in active pulsars that develop gaps around the closed-zone null surface. \citet{petri_global_2002} showed that such gaps are expected to open at radii of tens of stellar radii, well inside the light cylinder, and such a solution is demonstrated in Appendix~\ref{sec:pp_run}. Therefore, we conclude that our results are broadly applicable to most nonmillisecond pulsars. 

There are several directions for future work. While this study focused mainly on aligned rotators for simplicity, real pulsars possess finite obliquity. Our short oblique simulations demonstrate the onset and early nonlinear evolution of the diocotron instability, but further work is required to characterize the long-term behavior and associated charge transport in oblique rotators.

As the magnetosphere fills, the superrotating region moves closer to the light cylinder. While the diocotron instability growth rate is suppressed by relativistic effects \citep{petri_relativistic_2007}, understanding the nonlinear evolution in this regime will be important for connecting the instability to persistent observational signatures.

Additional work is required to better understand the radial diffusion, specifically the nature of diocotron mode cross-coupling and stochasticity and the effect of field-aligned motion in the closed zone. 

Finally, copious pair production---absent in our simulations---is required to sustain magnetospheric currents and enable radio emission in active pulsars. A fully self-consistent treatment of the diocotron instability in an active pulsar requires resolving the hierarchy $R_\star\ll R_{\rm d}\ll R_{\rm LC}$, in which pair cascades near the stellar surface (and possibly near the light cylinder) sustain the global current system, while the diocotron instability develops at radii where pair production can no longer close gaps in the closed zone.
Investigating how diocotron modes operate embedded within a larger pair-producing magnetosphere will help connect this work to the paradigm of nearly force-free pulsar magnetospheres and their observational signatures. 
\begin{acknowledgments}
We thank Hayk Hakobyan, Ashley Bransgrove, Robbie Ewart, Jasmine Parsons, and Nick Loudas for insightful discussions. Simulations were performed using resources managed and supported by Princeton University's Research Computing. This research was supported by NSF through Multimessenger Plasma Physics Center (grant PHY-2206607) and by Simons Foundation (grant MP-SCMPS-0000147). A.S. thanks the Institute for Advanced Study for hospitality.  
\end{acknowledgments}

\appendix
\section{Modeling the Potential Perturbation}\label{appx:modeling_charge}
\setcounter{equation}{0}
In Equation~(\ref{eq:rho_model}), we model the diocotron charge distribution as 
\begin{equation}
    \rho(r,\theta,\phi)=\sum_m{{Q}_m}\,Y_{m}(\theta,\phi)\frac{\delta(r-R_{\rm d})}{r^2},
\end{equation}
where 
\begin{equation}
    {Q}_m\equiv\int dV \,\rho(r,\theta,\phi)\, Y_m^\star(\theta,\phi)
\end{equation}
measures the total amount of charge participating in a diocotron mode with number $m$. We shall use moments to approximate the electric potential from this charge distribution.  
First, we calculate the interior and exterior multipole moments from the diocotron charge distribution. The interior multipole moment is 
\begin{align}
    I_m&=\int d{\bf r^\prime}\,\frac{\rho({\bf r^\prime})}{(r^\prime)^{m+1}}\sqrt{\frac{4\pi}{2m+1}}Y^*_m(\theta,\phi)\\
    &=\sqrt{\frac{4\pi}{2m+1}}\frac{{Q}_m}{R_{\rm d}^{m+1}}.
\end{align}
Similarly, the exterior multipole moment is 
\begin{equation}
    E_m=\sqrt{\frac{4\pi}{2m+1}}{{Q}_m}R_{\rm d}^{m}.
\end{equation}
Taking the interior multipole moment inside $R_{\rm d}$ and the exterior multipole moment outside, we read off the potential as 
\begin{equation}
    \Phi_{{\rm d}}=\frac{4\pi}{R_{\rm d}}\sum_{m}\frac{1}{2m+1}Q_mY_m(\theta,\phi)
    \begin{cases}
        \left(\frac{r}{R_{\rm d}}\right)^m & r<R_{\rm d}\\
         \left(\frac{R_{\rm d}}{r}\right)^{m+1} & r>R_{\rm d}
    \end{cases}
\end{equation}
For $\ell=m$ spherical harmonics, we have 
\begin{equation}
    Y_m=\frac{(-1)^m}{2^mm!}\sqrt{\frac{(2m+1)!}{4\pi}}\sin^m\theta \,e^{im\phi}
\end{equation}
so that 
\begin{equation}\label{eq:dioc_potential_appendix}
    \Phi_{{\rm d}}=2\frac{Q_c}{R_{\rm d}}\sum_{m>0}\frac{|Q_m|}{Q_c}\sin^m\theta \,\cos\varphi_m\,f(m)
    \begin{cases}
        \left(\frac{r}{R_{\rm d}}\right)^m & r<R_{\rm d}\\
         \left(\frac{R_{\rm d}}{r}\right)^{m+1} & r>R_{\rm d}
    \end{cases}
\end{equation}
where $\varphi_m\equiv m(\phi-\phi_m)$ and $\phi_m$ is the phase of the diocotron mode $m$, $Q_c\equiv (2\pi/3)\rho_{\rm GJ\star}R_\star^3$ \citep{Michel1991_book} is the pulsar central point charge, and, for brevity, 
\begin{equation}
    f(m)=\frac{1}{2^mm!}\sqrt{\frac{4\pi(2m)!}{2m+1}}
\end{equation}
is a real number that depends on $m$.

\section{Diffusive Radial Transport}\label{sec:diffusion_appendix}
\setcounter{equation}{0}
A particle in the equatorial plane experiences a radial $\bf E\times B$ drift velocity of $V_r=-cE_\phi/B_\theta$. At $r\gg R_{\rm d}$, $E_\phi$ is dominated by the azimuthal electric field of the $m=1$ diocotron mode. Taking the gradient of the $m=1$ potential perturbation (Equation~(\ref{eq:dioc_potential_appendix})) and evaluating at $\theta \approx \pi/2$, we obtain

\begin{equation}
    E_{\phi}=-\sqrt{\frac{8\pi}{3}}\frac{|Q_1|}{R_{\rm d}^2}\sin\varphi_1\left(\frac{R_{\rm d}}{r}\right)^3.
\end{equation}
In the equatorial plane, $B_\theta=B_\star(R_\star/r)^3$, giving
\begin{equation}
    V_{r}=V_{r0}\sin\varphi_1
\end{equation}
where 
\begin{align}
    V_{r0}&\equiv \sqrt{\frac{8\pi}{3}}\frac{|Q_1|}{R_{\rm d}^2}\left(\frac{r}{R_\star}\right)^3\frac{c}{B_\star}\\
    &=\frac{1}{3}\sqrt{\frac{8\pi}{3}}\frac{|Q_1|}{Q_c}\Omega_\star R_{\rm d} 
\end{align}
Here $\varphi_1 = \phi - \phi_1$, with $\phi$ the particle azimuthal coordinate and $\phi_1$ the phase of the $m=1$ mode. A particle not in resonance with the superrotating $m=1$ mode will undergo cyclic radial motion with zero net radial flux. However, as shown in Figures~\ref{fig:potential_vis}(a) and (b), both the $m=1$ mode strength and pattern speed fluctuate. Each particle can therefore complete each orbit at a slightly different radius than it started, resulting in radial diffusion. 
In Sections~\ref{sec:diff_mode_strength} and \ref{sec:diff_pspeed}, we calculate diffusion coefficients arising from fluctuations in the $m=1$ mode strength and pattern speed, respectively. Although we include both mechanisms for completeness, diffusion in our simulations is dominated by fluctuations in the mode strength.

Both higher-$m$ modes and cross-correlations between strength and pattern-speed fluctuations would increase the diffusion coefficient. Our results should therefore be regarded as a lower bound on radial transport driven by the diocotron instability in pulsar magnetospheres.
\subsection{Mode-strength Fluctuations}\label{sec:diff_mode_strength}

We first consider fluctuations in the $m=1$ mode strength, writing
$|Q_1|=\bar{Q}_1+\delta Q_1$, where $\bar{Q}_1$ is the average mode strength and $\delta Q_1$ is a stochastic fluctuation.
We assume that amplitude variations occur on timescales long compared to the pattern period. 

Applying the Taylor--Green--Kubo formula \citep[e.g.,][]{Chandrasekhar1943}, the diffusion coefficient is obtained by integrating the velocity autocorrelation function:
\begin{equation}D_{r,Q_1} = \int_0^\infty d\tau\, \langle V_r(t)V_r(t+\tau)\rangle . \end{equation}

 Recall that $V_r = \bar V_{r0}(1+\delta Q_1/\bar{Q}_1)\sin\varphi_1$. The zeroth-order term (in $\delta Q_1$) in the autocorrelation contributes to the diffusion only for resonant particles with $\Delta\bar{\omega}_1\approx 0$. In order to receive a coherent radial kick from the $m=1$ mode, the particle must perform most of an orbit in the pattern corotation frame before the mode decoheres and resonance is lost. However, the corotation frame orbital period $\sim 10 P_\star$ is much longer than the coherence time $\tau_{c,Q_1}\sim P_\star$. The rapid decoherence suppresses the effect of resonance on diffusion, which we therefore neglect.
 Likewise, the linear $\delta Q_1$ term in $\langle V_r^2\rangle$ vanishes since $\langle\delta Q_1\rangle=0$.
 Assuming that $\delta Q_1$ is uncorrelated with the orbital phase $\varphi_1$, we obtain
\begin{align}\label{eq:diff_integral}
 D_{r,Q_1}&=\frac{\bar{V}_{r0}^2}{2}\int_0^\infty{\rm d\tau}\,\left\langle\frac{\delta Q_1(t)\delta Q_1(t+\tau)}{\bar{Q}_1^2}\right\rangle\cos(\Delta\bar{\omega}_1\tau),
\end{align}
where $\Delta\bar{\omega}_1=\bar{\omega}_1-\Omega(r)$ is the {\it average} $m=1$ mode pattern frequency as seen in the frame rotating with the local azimuthal drift $\Omega(r)$. Assuming exponentially decorrelating fluctuations, the autocorrelation of $\delta Q_1$ is 
\begin{equation}
   \langle\delta Q_1(t)\delta Q_1(t+\tau)\rangle=\langle\delta Q_1^2\rangle e^{-\tau/\tau_{c,Q_1}},
\end{equation}
with correlation time $\tau_{c,Q_1}$ and $\langle \delta Q_1^2\rangle$ the variance of $Q_1$. The integral in Equation~(\ref{eq:diff_integral}) then evaluates to
 \begin{align}\label{eq:app_diffco_mstrength}
   D_{r,Q_1}&=\frac{\bar{V}_{r0}^2\tau_{c,Q_1}}{2}\left\langle\frac{\delta Q_1^2}{\bar{Q}_1^2}\right\rangle\frac{1}{1+\tau_{c,Q_1}^2\Delta\bar{\omega}_1^2}.
\end{align}

\subsection{Pattern-speed Fluctuations}\label{sec:diff_pspeed}
We now consider fluctuations in the pattern speed, writing
$\omega_1(t) = \bar{\omega}_1 + \delta\omega_1(t)$ while holding the mode
strength fixed. As in Section~\ref{sec:diff_mode_strength}, the relevant phase is
that sampled by a particle drifting at $\Omega(r)$, so the mean phase advances
at the Doppler-shifted rate $\Delta\bar{\omega}_1=\Omega(r)-\bar{\omega}_1$:
\begin{align}
    \varphi_1(t)
    &=\Delta\bar{\omega}_1t + \delta\varphi_1(t),
    \qquad
    \delta\varphi_1(t)\equiv\int_0^t{\rm d}t'\,\delta\omega_1(t').
\end{align}
Note that $|\delta\omega_1|$ is unaffected by the frame shift, so only the mean
frequency is Doppler shifted.
We again apply the Taylor--Green--Kubo formula for the diffusion coefficient.
After judicious application of the sine addition formula and dropping
oscillatory terms which average to zero, we obtain
\begin{align}
 D_{r,\omega_1}
&=\frac{\bar{V}_{r0}^2}{2} \int_0^\infty \mathrm{d}\tau\,
\cos\!\left[\Delta\bar{\omega}_1\tau\right]\left\langle
\cos[\delta\varphi_1(t+\tau)-\delta\varphi_1(t)]
\right\rangle\label{eq:coscos_int}
\end{align}
where $\delta\varphi_1$ is the stochastic phase perturbation.
Assuming Gaussian statistics for $\delta\omega_1$, we can use the cumulant
expansion to write
\begin{align}
    \left\langle
\cos[\delta\varphi_1(t+\tau)-\delta\varphi_1(t)]
\right\rangle
&=\exp\left[-\frac{1}{2}\int_t^{t+\tau}{\rm d}t'_1\int_t^{t+\tau}{\rm d}t'_2\,
\left\langle\delta\omega_1(t'_1)\delta\omega_1(t'_2)\right\rangle\right].
\label{eq:diff_doube_int}
\end{align}
For exponentially decorrelating noise, the autocorrelation function is
$\langle\delta\omega_1(t'_1)\delta\omega_1(t'_2)\rangle
=\langle\delta\omega_1^2\rangle\exp(-|t'_2-t'_1|/\tau_{c,\omega_1})$
for decorrelation time $\tau_{c,\omega_1}$.
Evaluating the double integral in Equation~(\ref{eq:diff_doube_int}), we obtain
\begin{align}
      \left\langle
\cos[\delta\varphi_1(t+\tau)-\delta\varphi_1(t)]
\right\rangle&=\exp\left[-\langle\delta\omega_1^2\rangle\tau_{c,\omega_1}^2
\left(\frac{\tau}{\tau_{c,\omega_1}}-1+\exp(-\tau/\tau_{c,\omega_1})\right)\right].
\end{align}
Plugging this into Equation~(\ref{eq:coscos_int}), we would now like to evaluate
 \begin{align}
      D_{r,\omega_1}&=\frac{\bar{V}_{r0}^2}{2} \int_0^\infty \mathrm{d}\tau\,
\cos\!\left[\Delta\bar{\omega}_1\tau\right]
\exp\left[-\langle\delta\omega_1^2\rangle\tau_{c,\omega_1}^2
\left(\frac{\tau}{\tau_{c,\omega_1}}-1+\exp(-\tau/\tau_{c,\omega_1})\right)\right].
\end{align}
This integral has no solution in terms of elementary functions. Its form is
controlled by the Kubo number $\eta\equiv\langle\delta\omega_1^2\rangle^{1/2}
\tau_{c,\omega_1}$, which compares the frequency excursion to the decorrelation
rate. For $\tau\gg\tau_{c,\omega_1}$ the exponent becomes linear in $\tau$, so
the phase correlation decays exponentially at the rate
\begin{equation}
    \Gamma \equiv \langle\delta\omega_1^2\rangle\,\tau_{c,\omega_1},
\end{equation}
i.e., the accumulated phase executes a random walk with diffusivity $\Gamma$.
The regime $\eta\ll1$ implies
$\Gamma\ll\tau_{c,\omega_1}^{-1}$, and this exponential decay governs the
integral at all frequencies of interest, giving
\begin{align}
    D_{r,\omega_1}
    &=\frac{\bar{V}_{r0}^2}{2}\,\frac{\Gamma}{\Gamma^2+\Delta\bar{\omega}_1^2}
    =\frac{\bar{V}_{r0}^2\,\tau_{\rm eff}}{2}
     \frac{1}{1+\Delta\bar{\omega}_1^2\tau_{\rm eff}^{2}},
    \qquad \tau_{\rm eff}\equiv\Gamma^{-1},
    \label{eq:app_diffco_pspeed}
\end{align}
which is a Lorentzian in the Doppler-shifted frequency of the same form as
Equation~(\ref{eq:app_diffco_mstrength}). The phase-decorrelation rate $\Gamma$
plays the role of $\tau_{c,Q_1}^{-1}$. Both channels therefore contribute
diffusion that is resonantly enhanced near the corotation radius
$\Omega(r_c)=\bar{\omega}_1$ and suppressed as
$\Delta\bar{\omega}_1^{-2}$ in the wings.

\section{Magnetosphere with Pair Production}\label{sec:pp_run}
\setcounter{equation}{0}
Active pulsars produce radio emission by generating electron--positron pair discharges over the polar cap \citep[e.g.,][]{philippov_origin_2020}. Pair production in such pulsars can fill some volume of the magnetosphere with plasma. \citet{petri_global_2002} calculated that this volume lies within $\sim 50R_\star$ in pulsars with $B_\star\sim10^{12}\,{\rm G}$. Outside $50 R_\star$, the magnetic field is too weak to support efficient $\gamma$--$B$ pair production. It is therefore important to understand the equilibrium structure of a magnetosphere featuring pair production with a radial cutoff, particularly whether it is susceptible to the diocotron instability.

We perform a 3D PIC simulation of an aligned pulsar magnetosphere with surface charge extraction and pair production. Similar to the method used by \citet{chen_filling_2020}, particles produce pairs within $R_{\rm cut}=3R_\star$ when they reach Lorentz factor $\gamma_{\rm thr}$. We use $\eta=\gamma_{\rm pc}/\gamma_{\rm thr}=32.5$, where $\gamma_{\rm pc}$ is the polar-cap potential drop. In this run, $R_{\rm LC}/R_\star=7.9$. In order to increase the polar-cap potential, the skin depth on the surface is reduced to $1.1$ cells. This is acceptable since the solution with pair production is not sensitive to small-scale electric fields on the surface. All other parameters are identical to the aligned pulsar simulation described in Section~\ref{sec:simulations}.

\begin{figure}[!htb]
    \centering
    \includegraphics[width=0.425\linewidth]{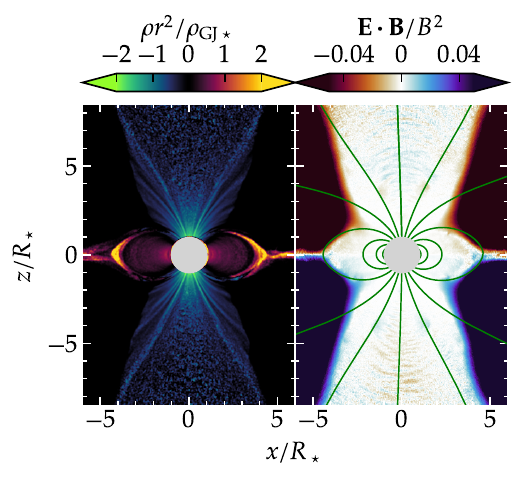}
    \caption{Slices in the $x$--$y$ plane of the pulsar with pair production enabled within $3R_\star$. Otherwise, the caption details are identical to those of Figure~\ref{fig:fig1_sideview}.}

    \label{fig:fig_sideview_pp}
\end{figure}

Figure~\ref{fig:fig_sideview_pp} shows the parallel electric field and charge density for this run with pair production. Inside $R_{\rm cut}$, the magnetosphere is approximately force-free and corotating. Outside, gaps open along the null surface, and we see a region of superrotating positrons in the closed zone. The solution is essentially scaled up with a new ``effective'' stellar radius of $R_{\rm cut}$. Due to the proximity of $R_{\rm cut}$ to $R_{\rm LC}$ in this scaled-down simulation, the differentially rotating part of the disk is close to the light cylinder and rotating at relativistic speeds. 

Relativistic effects suppress the evolution of the diocotron instability, and only a low-amplitude, high-$m$ mode develops \citep{petri_relativistic_2007}. Real pulsars with $P_\star\sim 1\,{\rm s}$, however, have an order of magnitude in scale separation between $R_{\rm cut}$ and $R_{\rm LC}$, providing ample room for the evolution of nonlinear diocotron modes in active pulsars with pair production.

\section{Screening of Diocotron Electric Fields}\label{sec:screening}
\setcounter{equation}{0}
As the diocotron electric field fluctuation rotates around the magnetosphere, it launches a low-frequency ($\omega=\Delta\omega_1\equiv\omega_1-\Omega_\star$ in the corotating frame) perturbation that propagates through the vacuum gaps toward the electron-filled domes. Where the perturbation meets a dome, the component of its electric field parallel to $\mathbf{B}$ is rapidly screened by the local charges, while the perpendicular component propagates into the dome as a fast magnetosonic wave, producing field-line displacements $\delta R\approx(2\pi c/\Delta\omega_1)(\delta E/B)$, as discussed in Section~\ref{sec:pc_effects}. The question of ``screening'' is therefore the question of how much of the wave electric field survives propagation through the dome to the polar cap.

A simple upper bound follows from noting that, in the highly magnetized limit, the charges can screen only the component of the fluctuation polarized along the local magnetic field. Since the perturbation of interest is $E_x$, the local screening fraction is the projection
\begin{equation}
    f=\hat{x}\cdot\hat{b}=\frac{3\sin\theta\cos\theta}{\sqrt{1+3\cos^{2}\theta}},
\end{equation}
evaluated at $\varphi=0$ for a dipole field. This expression reaches $f\rightarrow1$ at the null surface, $\theta\approx54.7^\circ$, where the field lines are horizontal and therefore aligned with $\hat{x}$: A dome extending that far could in principle screen the perturbation completely at its outer edge. The domes of interest here, however, are narrow columns standing above the poles, whose opening angle $\theta_{\rm dome}$ is small and decreases with height, so that $f\approx\tfrac{3}{2}\theta_{\rm dome}\ll1$ over most of the column. For $\theta_{\rm dome}=20^\circ$ the bound gives $f\approx0.5$, so at most half of the amplitude of the $E_x$ perturbation can be screened; on the axis itself, $f=0$, and the perturbation is transmitted without screening at all.

This is an upper bound because $f$ is evaluated at the location most favorable to screening: at $\varphi=0$, where the field lines lie in the $x$--$z$ plane, and at the outer edge of the dome, where $\theta$ is largest. In reality, the rotation of the field lines away from $\hat{x}$ in azimuth and toward $\hat{z}$ with decreasing colatitude both reduce the projection further.
Because the diocotron fluctuation can be damped only where it is polarized along the magnetic field, the field-line geometry strongly suppresses magnetospheric screening near the axis.

\bibliography{pulsars}
\bibliographystyle{aasjournalv7.1}

\end{document}